\documentclass[journal=jpcbfk,manuscript=article,layout=onecolumn,articletitle=false,super=true]{achemso}
\setkeys{acs}{articletitle=true}
\setkeys{acs}{keywords = true}
\setkeys{acs}{etalmode=truncate,maxauthors=10}
\usepackage{graphics}
\usepackage{graphicx}
\usepackage{epsfig}
\usepackage{amssymb}
\usepackage{amsthm}
\usepackage{rotating}
\usepackage{mathtools}
\usepackage{hyperref}
\usepackage{chemfig}

\newcommand{\fref}[1]{Figure~\ref{#1}}

\author{Chandan Kumar Das}
\affiliation {Department of Chemistry, Indian Institute Of Technology Kanpur, Kanpur 208016, India}
\author{Nisanth N. Nair}
\affiliation {Department of Chemistry, Indian Institute Of Technology Kanpur, Kanpur 208016, India}
\email{nnair@iitk.ac.in}

\title{Hydrolysis of Cephalexin and Meropenem by New Delhi Metallo $\beta$--Lactamase: Substrate Protonation Mechanism is Drug Dependent}

\begin{document}






\begin{figure*}
\begin{center}
\includegraphics[scale=1.0]{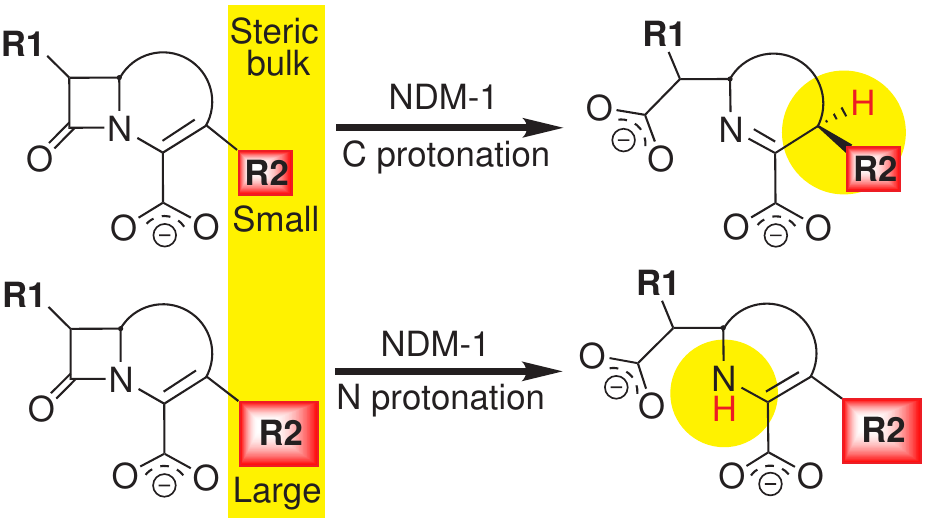}
\end{center}
\end{figure*}

%
\begin{abstract}
Emergence of antibiotic resistance due to New Delhi Metallo $\beta$-lactamase 
(NDM-1) bacterial enzymes is of great concern due to their ability to hydrolyze wide range of 
antibiotics.
Efforts are ongoing to obtain the atomistic details of the hydrolysis 
mechanism in order to develop novel drugs and inhibitors against NDM-1.
Especially, it remains elusive how 
drug molecules of different family of antibiotics 
are hydrolyzed by 
NDM-1 in an efficient manner.
Here we report the detailed molecular mechanism of NDM-1 catalyzed hydrolysis of cephalexin, a 
cephalosporin family drug, and meropenem, a carbapenem family drug.
This study employs molecular dynamics (MD) simulations using hybrid                 
 quantum mechanical/molecular mechanical 
(QM/MM) methods at the density functional theory level, 
based on which reaction pathways and the associated free energies are 
obtained.
We find that the mechanism and the free energy barrier for the ring--opening 
step are the same for both the drug molecules, while the subsequent protonation 
step differs.
In particular, we observe that the mechanism of the protonation step depends 
on the {R2} group of the drug molecule.
Our simulations show that allylic carbon protonation occurs in the case of cephalexin 
drug molecule where Lys211 is the proton donor and the proton transfer occurs via a water chain 
formed (only) at the ring--opened 
intermediate structure.
Based on the free energy profiles,  the overall kinetics of the drug hydrolysis is  discussed. 
Finally, we show that the proposed mechanisms and free energy profiles could explain various experimental 
observations.
\end{abstract}
%
%
%
%
\section{Introduction}
Escalating antimicrobial resistance (AMR) has been realized as a global threat to the public health in the recent times.~\cite{WHO_REPORT_2014}
AMR related death is estimated to be about 700~000 per year and it is predicted to be the biggest devastating problem by 2050
with 10 million deaths per year.~\cite{AMR_REPORT_2016}
Implementation of effective measures to combat drug resistance shown by bacteria, in particular, development of 
novel antimicrobial drugs, has to be expedited.
One of the ways in which bacteria develop antibiotic resistance is by expressing $\beta$--lactamase enzymes,
which catalyze the hydrolysis of antibiotic drug molecules in an efficient way.
New Delhi Metallo-$\beta$-lactamase-1 (NDM-1) is one such enzymes which was identified only in 2009,\cite{Yong_2009_AAC} however  becoming
widely expressed in bacteria causing hospital--acquired and community--acquired infections.\cite{Kumarasamy_2010_LID}
Of great concern, NDM-1 carrying bacteria are known to hydrolyze nearly all the  
clinically used $\beta$--lactam antibiotics including penicillins, cephalosporins and even carbapenems considered 
as the last resort drugs.\cite{Nordmann_2011_TM,Mundy:2013,Tada_2014_AAC,Shrestha_2015_AAC} 
Thus there is an urgent need to develop effective inhibitors for NDM-1 so that 
life threatening infections due to NDM-1 carrying ``superbugs'' be cured.

Development of inhibitors is majorly aided by the knowledge of structure and reactivity of the active site. 
Over the last few decades a number of crystallographic,
\cite{Feng_2014_JACS, Zhang_2011_FASEB, Kim_2013_FASEB, Dustin_2012_JACS}
spectroscopic,\cite{Wang_1998_JACS, Rydzik_2014_ANGEW, Yamaguchi_2005_JBC, Llarrull_2007_JBC, Rasia_2004_JBC, Mariana_2008_JACS, Breece_2009_JACS, Hawk_2009_JACS, Silvia_1999_BIOCHEMISTRY, Zhenxin_2008_JACS, Yang_2012_BIOCHEMISTRY, Yang_2014_JACS} 
and computational\cite{Kim_2013_FASEB, Zheng_2013_JPCB, Matteo_2007_JACS, Hwangseo_2005_JACS, Xu_2007_JACS, Xu_2007_JPCA, Zhu_2013_JCAMD, Dimas_2002_BIOCHEMISTRY, Ravi_2015_ACSCAT} 
studies have reported which shed light on the molecular structure and 
catalytic reactions of di--Zn metallo $\beta-$lactamases (MBLs), including NDM-1.
Active site of NDM-1 has two Zn(II) binding sites bridged with a hydroxide (W1). 
The Zn1 site is coordinated with His120, His122, and His189, while the Zn2 site is bound to Asp124, Cys208, and His250. 
The existence of a water molecule bound to Zn2 in the apoenzyme is also debated.\cite{Meini_2015_FEBS}

\begin{figure}[t]
\begin{center}
\includegraphics[scale=1.0]{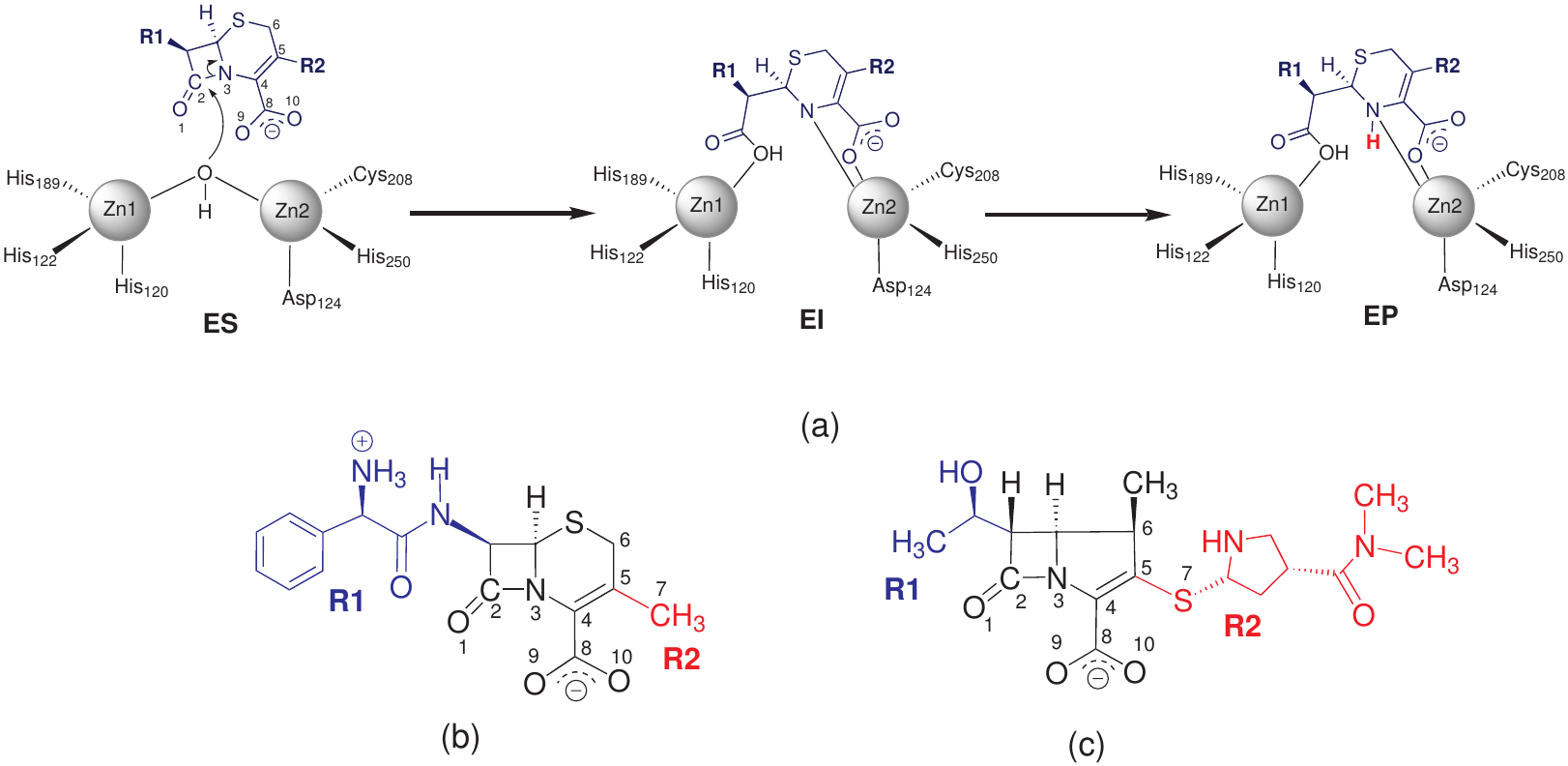}
\end{center}
\caption[]{ (a) General mechanism of NDM-1 catalyzed $\beta$--lactam antibiotic hydrolysis. Chemical structure of cephalexin (b) and meropenem (c).  {R1} and R2 groups are shown in blue and in red, respectively. \label{general-mech-drugs} }
\end{figure}

Based on the existing studies on NDM-1 and other di--Zn Class B1 MBLs,\cite{Mugesh:12,Palzkill:13,Meini_2015_FEBS} 
the hydrolysis is initialized by the nucleophilic attack ({\bf ES} $\rightarrow$ {\bf EI}) where the 
bridging hydroxide attacks the $\beta$--lactam ring of the drug molecules (see \fref{general-mech-drugs}). 
This leads to the intermediate structure {\bf EI} where the $\beta$--lactam N of the ring--opened drug molecule is 
bound to Zn2.
The subsequent step involves a proton transfer to N$_3$ (or C$_5$) and is followed by 
decoordination of the drug from the active site, completing the hydrolysis; see \fref{general-mech-drugs} for the atom numbers followed.
However, the detailed mechanism of hydrolysis is not fully understood.
Mostly, the proton donor has to be identified and the mechanism of proton transfer has to be ascertained.
Further, the rate--determining step of the whole process has to be established.
%

In general, proton transfer to N$_3$ is anticipated,
while C$_5$ protonation has been also noticed in some experimental studies.~\cite{Mariana_2008_JACS,Feng_2014_JACS}
A tautomerization where C$_4$--C$_5$ double bond is migrated to N$_3$--C$_4$ could result in
protonation of C$_5$ instead of N$_3$, and was noted in the experiments by Tioni~et~al.~\cite{Mariana_2008_JACS} 
while studying the carbapenem hydrolysis by BcII MBL.
Protonation of C$_5$ protonation was observed in the crystal structure of 
ring--opened cephalexin bound to NDM-1.~\cite{Feng_2014_JACS}
Thus, both N$_3$ and C$_5$ are identified as the likely proton acceptors during 
{\bf EI} $\rightarrow$ {\bf EP} reaction.
The factors controlling the
protonation mechanism are yet to be scrutinized.

%
Asp124 bound to Zn2 was proposed to get protonated during the nucleophilic attack
of the bridging hydroxyl group in NDM-1~\cite{Zheng_2013_JPCB,Ravi_2015_ACSCAT} and other MBLs.~\cite{Xu_2007_JACS,Hwangseo_2005_JACS}
Thus the protonated Asp124 formed after the nucleophilic attack could act as the proton donor.\cite{Jonathan_2007_BIOCHEMISTRY} 
However, detailed mutational studies indicate that Asp124 is not likely to be involved in the protonation
of the intermediate.\cite{Crowder:2004,Llarrull_2007_JBC,Jonathan_2007_BIOCHEMISTRY} 
The same conclusion was also arrived by the QM/MM MD study of meropenem hydrolysis by NDM-1.~\cite{Ravi_2015_ACSCAT}
The bridging hydroxyl group was proposed as the proton donor by Zhang and Hao based on the
crystal structure of ampicillin complexed with NDM-1.\cite{Zhang_2011_FASEB}
Observation of water or hydroxyl group bound to the Zn ions in the crystal structures of the reaction 
intermediates~\cite{Dustin_2012_JACS} also hints that protonation might occur from the dissociation of the bulk water 
coordinating to the Zn site during the hydrolysis reaction.
This mechanism was also supported by the previous QM/MM MD simulation, 
where a bulk water coordinates to Zn1 in the {\bf EI} structure, which subsequently 
dissociate to protonate N$_3$.~\cite{Ravi_2015_ACSCAT}

Formation of anionic intermediate is also reported based on various spectroscopic data for di--Zn 
MBLs\cite{Wang_1998_JACS,Silvia_1999_BIOCHEMISTRY,Wang_1999_Biochemistry,Mariana_2008_JACS,Yang_2012_BIOCHEMISTRY,Yang_2014_JACS}
and further supported by QM/MM simulations.\cite{Hwangseo_2005_JACS,Zhu_2013_JCAMD,Xu_2007_JACS,Zheng_2013_JPCB,Ravi_2015_ACSCAT}
X--ray structures of di--Zn MBLs with ring--opened drug molecules have 
 N$_3$--Zn2 coordination, thus these structures are likely to be the ``trapped" {\bf EI} intermediate state.~\cite{Dustin_2012_JACS,Zhang_2011_FASEB,Gamblin:2005,Mariana_2008_JACS,Feng_2014_JACS,Dieberg:2005}
Spectroscopic and kinetic data of NDM-1 catalyzed hydrolysis of chromacef and nitrocefin
suggest that the rate constant for {\bf EI} $\rightarrow$ {\bf EP} ($k_3$) is nearly 10 times smaller than {\bf ES} $\rightarrow$ {\bf EI} ($k_2$)
hinting on the accumulation of the anionic intermediate.~\cite{Yang_2012_BIOCHEMISTRY,Yang_2014_JACS}
Based on the kinetic parameters derived from these experiments, it was concluded that the rate determining step
is the protonation of the anionic intermediate (i.e.  {\bf EI} $\rightarrow$ {\bf EP} ).
This is also consistent with the kinetic studies where solvent isotope effect was observed in the hydrolysis of various drug molecules
by di--Zn MBLs.\cite{Bounaga:98,Llarrull_2007_JBC,Wang_1998_JACS} 
Previous QM/MM study also agrees with the formation of stable anionic intermediate, where the negative charge of $\beta$--lactam N
is stabilized by Zn2.\cite{Ravi_2015_ACSCAT}
%
%

The predicted mechanism from QM/MM MD simulation is composed of a water diffusion towards the active site after the formation of the {\bf EI} complex.\cite{Ravi_2015_ACSCAT}
The diffused water molecule coordinates with Zn1, which subsequently acts as the proton donor.
However, coordination of a bulk water molecule is anticipated to be a slow process in the complexed structure, 
and thus an alternative mechanism was proposed  for di--Zn MBLs  by Vila and co--workers where the apoenzyme active site 
is considered to be composed of Zn2 coordinated with a additional water molecule (W2).~\cite{Meini_2015_FEBS}
According to this mechanism, the proton transfer to the anionic intermediate occurs directly from W2.
Klein and co--workers have carried out QM/MM MD simulation of CCrA catalyzed hydrolysis using this active site model
and found that
accumulation of anionic species doesn't occur
since proton transfer to N$_3$ occurs spontaneous 
with the formation of {\bf EI}.~\cite{Matteo_2007_JACS}
Based on this a pre--coordinated water model can excluded as accumulation of anionic species is 
observed in various experiments. 
Further, crystallographic structures was found without any bulk water molecules coordinated at the Zn sites in the trapped reaction intermediate states.\cite{Dustin_2012_JACS}

In this work,  we investigate various mechanistic routes  
for cephalexin and meropenem hydrolysis catalyzed by NDM-1.
Primarily, we address the question of N$_3$ protonation versus C$_5$ protonation.
%
%
First principle density functional theory (DFT) based QM/MM  Car-Parrinello MD~\cite{Laio_2002_JCP} combined with conventional~\cite{Laio_2002_PNAS,Laio_2003_PRL}
 and the recently developed well--sliced metadynamics\cite{Shalini_2016_JCC} approaches
are used here for simulating chemical reactions and to obtain free energy barriers.

\section{Methods and Models}
We followed a three step simulation strategy in this work. 
At first, we carried out MM force--field MD simulations of the fully solvated protein--drug complexes {\bf ES} of
cephalexin and meropenem.
This was followed by DFT based QM/MM MD simulations to further equilibrate these systems.
Finally, taking the equilibrated structures, we performed QM/MM metadynamics or well--sliced metadynamics simulations to model chemical reactions.
The reaction mechanism and free energy barriers of all  elementary steps are obtained from these simulations.
Henry--Michaelis complexes of NDM-1 with cephalexin and meropenem were built from the crystal structures of NDM-1 complex with hydrolyzed drug molecules,
 PDB IDs 4RL2~\cite{Feng_2014_JACS} and 
 4EYL~\cite{Dustin_2012_JACS}, respectively. 
Restrained Electrostatic Potential (RESP) derived point charges of the drug molecules were computed using 
the R.E.D tools~\cite{redtools} (see SI) and the potentials for the drug molecules were described 
by the GAFF force--field.~\cite{gaff} 
The protonation states for all the ionizable residues of the protein were set to that 
correspond to pH$=7$. 
We considered Asp124 in the deprotonated form, as found in our earlier study.~\cite{Ravi_thesis} 
All the crystal structure water molecules were retained while building the initial structure.
Two Zn atoms are bridged by a hydroxyl, and  
the Zn1 site is coordinated with His120, His122, and His189, 
whereas  the Zn2 site is bound to Asp124, Cys208, and His250;
see SI~Figure~S1. 
Zn--ligand coordination was modeled following the work of Suarez~{\em et~al}.~\cite{Dimas_2002_BIOCHEMISTRY}
While modeling the cephalexin--NDM-1 complex three Na$^+$ ions were added to 
neutralize the whole system which was solvated with 8914 water molecules in a periodic 
box of dimensions 66$\times$75$\times$76~\AA$^3$. 
In order to model the solvated meropenem--NDM-1 complex, we added four Na$^+$ ions to the system containing the protein 
in a periodic box of 70$\times$68$\times$72~\AA$^3$ containing 7990 water molecules.
The parm99~\cite{parm99} version of the AMBER force field was used to model the protein.  
MM water molecules were treated using the TIP3P~\cite{tip3p} 
force field.
The particle-mesh Ewald method~\cite{pme} was employed to compute the long range electrostatics, and
a non-bonded interaction cutoff of 15~{\AA} was used.
Sander module of the AMBER program package was used to perform MM MD simulations.\cite{amber} 
These MD simulations were carried out with a  time step of 1~fs.
All the atoms in the system, including the solvent molecules were relaxed during the MD simulations.
We performed nearly 2~ns $NPT$ simulation at 1~atm and 300~K using Berendsen barostat~\cite{Berendsen_1984_JCP} 
and Langevin thermostat~\cite{Loncharich_1992_BIOPOLYMERS} in order to obtain the equilibrium density.
With the equilibrated cells, we carried out 10~ns of $NVT$ MD simulations.
These simulations were carried out till the RMSD fluctuations of the backbone and the active sites 
 have converged satisfactorily; see SI~Figure~S3.
%

Hybrid QM/MM simulations~\cite{Warshel_1976_JMB,Carloni:2002,Laio_2002_JCP,Dominik_AIMD_book} 
of the equilibrated structures obtained from the MM simulations were performed using 
the CPMD/GROMOS interface, as available in the CPMD package.~\cite{cpmdpackage} 
In these simulations, the whole  
drug molecule, two Zn ions, the embedded hydroxide and the 
coordinated side chains of His120, His122, Asp124, His189, Cys208, His250, were  
treated quantum mechanically; see SI~Figure~S2.
Additionally, one water molecule (W2) was also treated quantum mechanically when modeling {\bf EI1} $\rightarrow$ {\bf EI2} 
reaction, while the side chain of Lys211 as well as two water molecules (W3 and W4) were described quantum mechanically for
modeling  {\bf EI1} $\rightarrow$ {\bf EI2$^\prime$} reaction; see SI~Figure~S4(b).
We used capping hydrogen atoms to saturate the dangling chemical bonds of the QM part.
Capping hydrogen atoms were placed between the C$_\beta$--C$_\gamma$ bonds in His120, His122, His189 and His250,  
between the C$_\alpha$--C$_\beta$ bonds in Asp124 and Cys208 
and between C$_ \delta$--C$_ \epsilon$ bond in Lys211 (SI~Figure~S2).
A cubic QM box with a side length of 27.5~{\AA} was used, and 
the QM subsystems were treated at the DFT level using the PBE~\cite{Perdew_1992_PRB} density functional
and ultra soft pseudopotentials were used to describe the core--potentials.~\cite{Vanderbilt_1990_PRB}  
Plane wave basis set with a plane wave cutoff of 
25~Ry was used in these calculations. 
The QM/MM electrostatic coupling was treated using the scheme by Laio~{\em et~al}.~\cite{Laio_2002_JCP}
The QM/MM MD runs were carried out using the Car-Parrinello method with a time step of 
0.125~fs and a fictitious mass of 700~a.u. for the orbital degrees of freedom.~\cite{Car_1985_PRL, Dominik_AIMD_book}
The nuclear and the orbital degrees of freedom were thermostated using 
Nos{\'e}-Hoover chains thermostats.~\cite{Nose_1992_JCP}
Ionic temperature in the MD simulations was set to 300~K.
We carried out $NVT$ QM/MM MD simulations for nearly 10~ps for both the systems, before starting metadynamics simulations.

In order to accelerate the sampling of chemical reactions, thus to obtain reaction mechanism and free energy barriers,
we used the metadynamics method.~\cite{Laio_2002_PNAS,Laio_2008_RPP, Bernd_2006_ACR, Barducci_2011_CMS,Valsson_2016_ARPC} 
By applying dynamically grown biasing potentials, metadynamics technique enhances the sampling of a selected set of collective variables (CVs) that are relevant for a chemical reaction. 
We employed the extended Lagrangian variant of metadynamics~\cite{Laio_2003_PRL}, and the details of the metadynamics
simulation setup and the choice of CVs for various elementary steps of the hydrolysis reaction can be found in SI~Section~8. 
We also used the well--sliced metadynamics approach,~\cite{Shalini_2016_JCC} for exploring broad and unbound free energy surfaces 
 by a combination of umbrella sampling and metadynamics techniques.
Initial structures of metadynamics simulations were taken from the equilibrated structures obtained from QM/MM $NVT$ runs.

The transition state for the step {\bf ES}$\rightarrow${\bf EI} was 
obtained through committer analysis~\cite{Christoph_TPS} ; see Section~SI~10.
Accuracy of the free energy estimates, considering the typical errors due to metadynamics and the density functional is about 
($W$+1)~kcal~mol$^{-1}$,  
where $W$ is the height of the Gaussian potentials used in metadynamics; see Section~SI~11.

\section{Results and Discussion}
%
\begin{figure*}
\begin{center}
\includegraphics[scale=0.7]{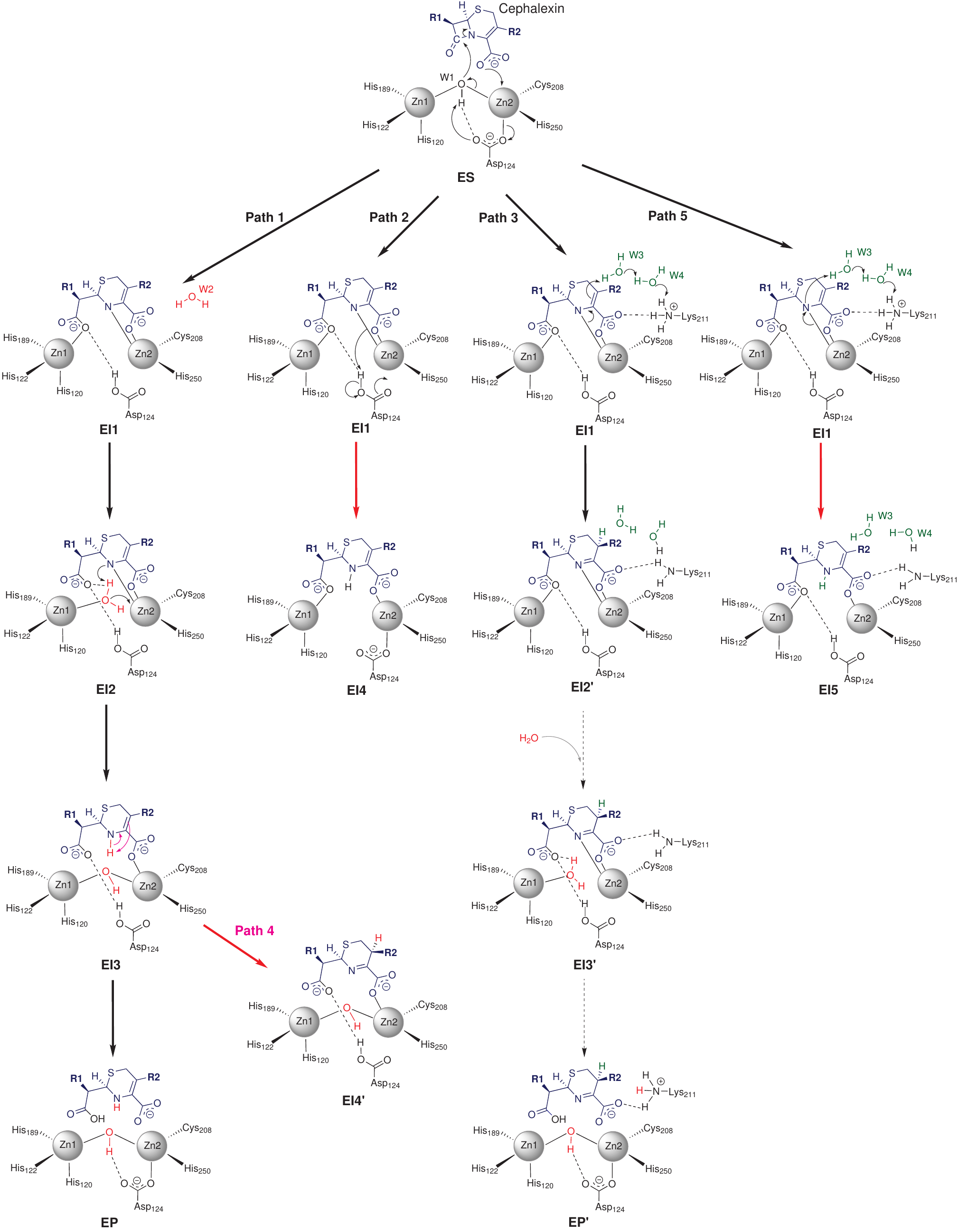}
\end{center}
\caption[]{Mechanistic paths of cephalexin hydrolysis by NDM-1 explored in our QM/MM--simulations. 
Red arrows indicate energetically unfavorable reaction pathways.
Dotted arrows are hypothesized steps on the basis of our simulation. 
See SI~Figure~S23 for the mechanistic routes investigated for meropenem hydrolysis. \label{fig:mechanism:all} }
\end{figure*}
\begin{figure*}
\begin{center}
\includegraphics[scale=0.4]{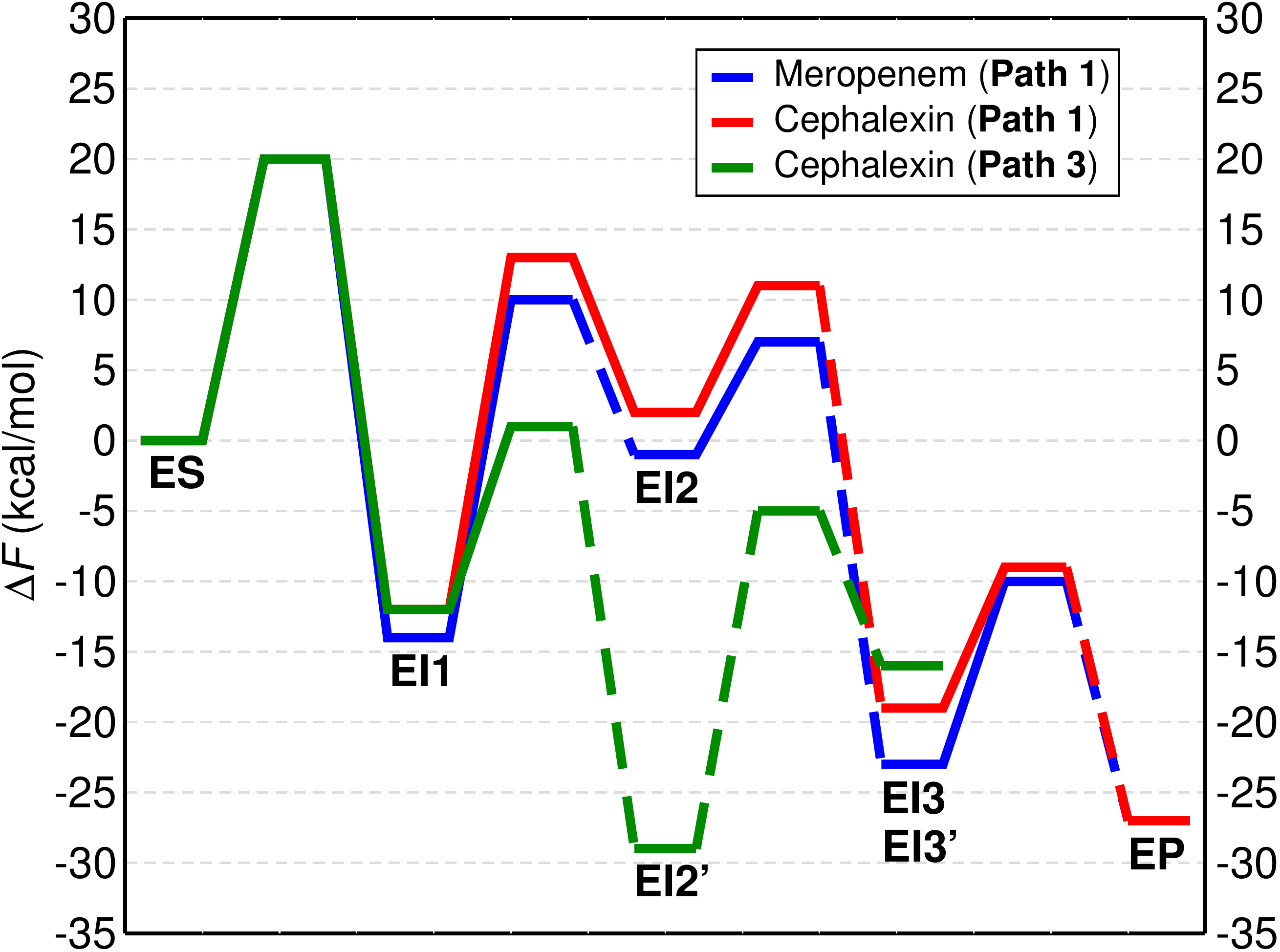}
\end{center}
\caption[]{Free energy profile for the various pathways for meropenem and cephalexin hydrolysis. Solid lines are computed from our calculations while dotted lines are tentative and not determined from our computations. \label{fig:FEProfile}}
\end{figure*}
\begin{figure*}
\begin{center}
\includegraphics[scale=0.25]{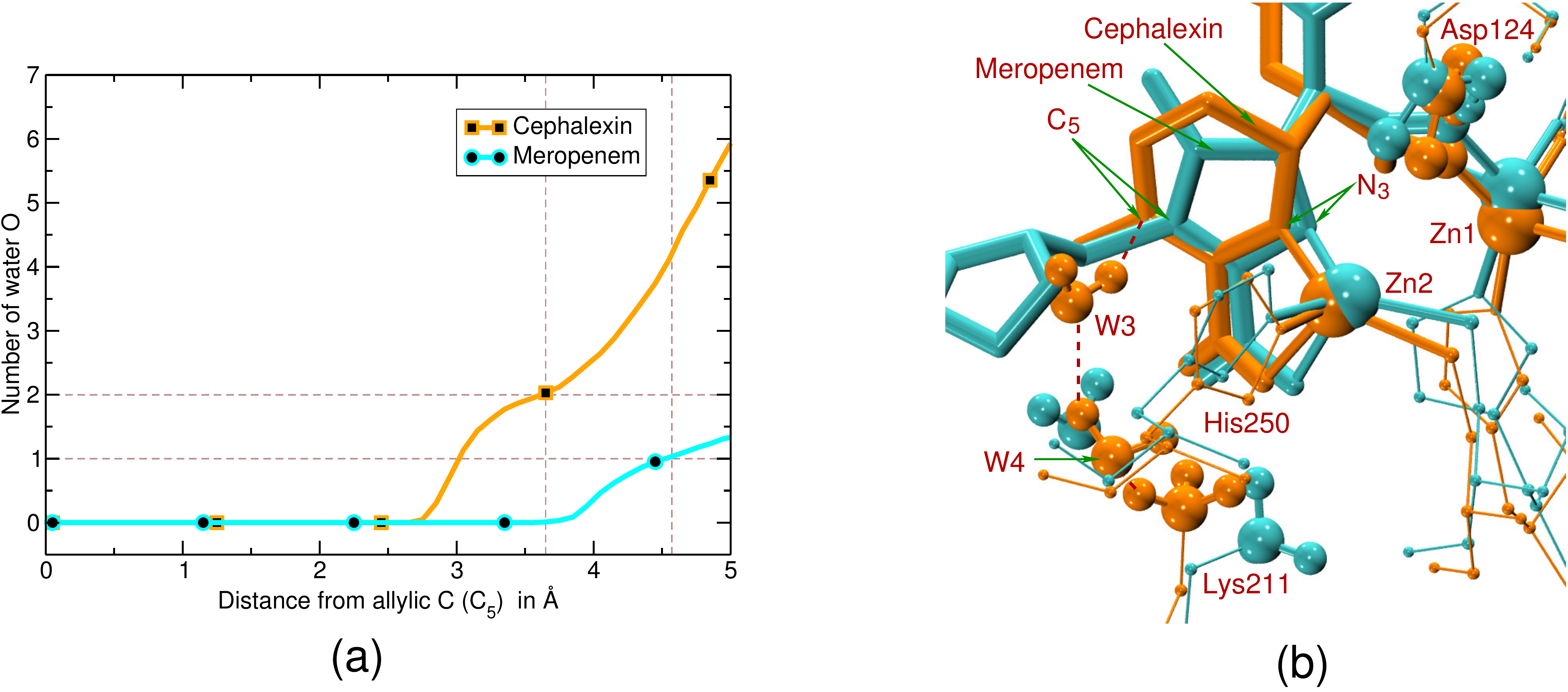}
\end{center}
\caption[]{ (a) The number of water molecules surrounding to the allylic C of $\beta$--lactam N (C$_5$) during $NVT$ QM/MM simulation of the {\bf EI1} is calculated by integrating radial distribution function ({\it g}({\it r})) of all water oxygen atoms from C$_5$. 
(b) Overlapped {\bf EI1} structures after QM/MM equilibration for the case of meropenem (cyan) and cephalexin (orange). \label{int:gr} }
\end{figure*}
%
%
In order to study the hydrolysis of cephalexin, first we equilibrated the {\bf ES} structure
using MM, QM/MM MD simulations (see SI~Figure~S1 and SI~Section~4) and then we explored five different reaction pathways using 
a series of metadynamics simulations (see SI~Table~S3).
%

The {\bf Path 1}, which we investigated first, is 
{\bf ES} $\rightarrow$ {\bf EI1} $\rightarrow$ {\bf EI2} $\rightarrow$ {\bf EI3} $\rightarrow$ {\bf EP}; see \fref{fig:mechanism:all}. 
This pathway is identical to that we proposed in our earlier study
on meropenem hydrolysis.~\cite{Ravi_2015_ACSCAT}
The elementary reaction steps were explored using independent metadynamics 
simulations, and based on which, free energy barriers for the elementary steps were computed; see \fref{fig:FEProfile}.
The first step {\bf ES} $\rightarrow$ {\bf EI1} involves the ring--opening reaction where
the bridging O$_{\rm W1}$ attacks C$_2$.
Most importantly, in the {\bf ES} structure, Zn1 and Zn2 are only weakly coordinated to O$_1$ and O$_9$, 
respectively, while these interactions strengthen as the O$_{\rm W1}$$\cdots$Zn2 coordination interaction is weakened.
Hydrogen bonding interaction between H$_{\rm W1}$ and Asp124:O$_{\delta 1}$
is favorably orienting W1 for the nucleophilic attack on C$_2$. 
Subsequently, O$_{\rm W1}$ attacks C$_2$, and C$_2$--N$_3$ bond is cleaved.
Almost simultaneous with C$_2$--N$_3$ bond breakage, we observed coordination of N$_3$ to Zn2, 
Zn2--Asp124:O$_{\delta 2}$ bond dissociation, and H$_{\rm W1}$ transfer from O$_{\rm W1}$ to 
Asp124:O$_{\delta 1}$; see SI~Figure~S20.
It is interesting to note that these processes were occurring almost simultaneously near the transition state. 
For further verification we characterized the transition state for {\bf ES} $\rightarrow$ {\bf EI1} reaction
through committor analysis and the structure of the transition state is shown in SI~Figure~S8(a). 
In the transition state, C$_2$--N$_3$ distance is 1.58~{\AA}, and O$_{\rm W1}$--C$_2$ distance is 1.51~{\AA}.
Of great importance, we notice that H$_{\rm W1}$ is delocalized between O$_{\rm W1}$ and Asp124:O$_{\delta 1}$ in the transition state structure. 
%
%
%
%
%
According to the computed free energy profile in \fref{fig:FEProfile}, we find that 
the free energy barrier for {\bf ES}$\rightarrow${\bf EI1} (20~kcal~mol$^{-1}$; see SI~Figure~S9) was found to be 
smaller than the reverse step {\bf EI1}$\rightarrow${\bf ES} (32~kcal~mol$^{-1}$; see SI~Figure~S10)

In the {\bf EI1} structure, a bulk water molecule (W2) is located near 
the hydrophilic pocket consisting of the C$_4$ carboxylate, Lys211, and backbone amide hydrogen of Asn220; see SI~Figure~S4(b).
{\bf EI1} $\rightarrow$ {\bf EI2} composes of diffusion of W2 
(which was treated quantum mechanically in this simulation), 
followed by coordination to Zn1 and the associated decoordination of O$_1$ and O$_{\rm W1}$ from Zn1; see SI~Figure~S5. 
The free energy barriers for {\bf EI1} $\rightarrow$ {\bf EI2} and for the reverse step were computed as 
25~kcal~mol$^{-1}$ (SI~Figure~S11) and 11~kcal~mol$^{-1}$ (SI~Figure~S12), respectively.

{\bf EI2} $\rightarrow$ {\bf EI3} is the N$_3$ protonation step.
While modeling this step, we sampled two protonation pathways in the same simulation: 
(a) protonation by the transfer of H$_{\rm W1}$ protons, which are coordinated to 
Asp124:O$_{\delta 1}$ and O$_{\rm W1}$; 
(b) protonation by the transfer of H$_{\rm W2}$ protons of the
W2 water molecule coordinated to Zn1.
However, we have found that transfer of H$_{\rm W2}$ to N$_3$ is the preferred route for N$_3$ protonation. 
The free energy barrier for this reaction was computed as 9~kcal~mol$^{-1}$ (SI~Figure~S13).
With the formation of {\bf EI3}, the hydroxyl bridge is reformed as in the {\bf ES} structure, yet
Asp124:O$_{\delta 1}$ is protonated and Asp124:O$_{\delta 2}$ 
is weakly bound to the Zn2 site.
Thus we modeled {\bf EI3} $\rightarrow$ {\bf EP}, 
where we sampled the coordination of Asp124:O$_{\delta 2}$ and decoordination of O$_9$ from 
the Zn2 site. 
Here we observed the product ({\bf EP}) formation, 
where the H$_{\rm {W1}}$ is first transferred to O$_{\rm {W1}}$ from Asp124:O$_{\delta 1}$, 
and Asp124:O$_{\delta 2}$ coordinates to Zn2. 
This process is also synchronous with the decoordination of O$_9$ from Zn2.
The free energy barrier for this reaction was computed as 10~kcal~mol$^{-1}$ (SI~Figure~S14).

Having determined the mechanism and free energy profile along {\bf Path 1}, we
modeled the reaction along {\bf Path 2}, where the reaction proceeds through
{\bf ES} $\rightarrow$ {\bf EI1} $\rightarrow$ {\bf EI4}; see \fref{fig:mechanism:all}.
Here, we explored the possibility of protonation of N$_3$ in {\bf EI1}, before the water diffusion to
the active site,
through a direct proton transfer from W1 to N$_3$. 
This is motivated by the work of Zhang and Hao, who proposed this mechanism 
based on the close distance between O$_{\rm W1}$ and N$_3$ in the crystal structure 
PDB ID 3Q6X.~\cite{Zhang_2011_FASEB}
In our simulation, we observed the proton transfer to N$_3$ from Asp124:O$_{\delta 1}$. 
The free energy barrier for this reaction was computed to be 36~kcal~mol$^{-1}$ (SI~Figure~S15).  
Thus it is clear that {\bf Path 1} is preferred over {\bf Path 2}; see \fref{fig:FEProfile}).

Subsequently, we explored {\bf ES} $\rightarrow$ {\bf EI1} $\rightarrow$ {\bf EI2$^\prime$} $\rightarrow$ {\bf EI3$^\prime$} $\rightarrow${\bf EP},  i.~e. {\bf Path 3}; see \fref{fig:mechanism:all}.
Along this pathway, we investigated the possibility of C$_5$ protonation.
On our careful analysis of the equilibrated {\bf EI1} structure, we observed that a water molecule (W3) is in the vicinity of C$_5$.
Interestingly, W3 is hydrogen bonded to another water molecule (W4), which in turn is hydrogen bonded to protonated Lys211; see SI~Figure~S4(b).
Thus we explored proton transfer to C$_5$ from Lys211 through W3 and W4.
Tautomerization of the double bond from C$_4$--C$_5$ to C$_4$--N$_3$ in the six membered ring should also take place during this reaction.
We successfully simulated this reaction, and found that the free energy barrier for {\bf EI1} $\rightarrow$ {\bf EI2$^\prime$} is  13~kcal~mol$^{-1}$ (SI~Figures~S16 and S22). 
Free energy profiles for {\bf Path 1} and {\bf Path 3} (\fref{fig:FEProfile}) indicate that cephalexin hydrolysis  proceeds preferably through C$_5$ protonation along {\bf Path 3}, rather than via N$_3$ protonation along {\bf Path 1}, since the former has a smaller free energy barrier for protonation.
We hypothesize that, {\bf EI2$^\prime$}  further results in {\bf EP$^\prime$} after the coordination of a bulk water molecule to the active site as in {\bf EI1} $\rightarrow$ {\bf EI2} along {\bf Path 1}.
The crystal structures of the hydrolyzed cephalexin bound to NDM-1 reported by Feng~{\em et al.}~\cite{Feng_2014_JACS} where C$_5$ is in tetrahedral coordination also embeds an oxygen atom between the two Zn ions.
%
On the other hand, the observation of C$_5$ protonation in Tioni~et~al.~\cite{Mariana_2008_JACS} experiment is anticipated to 
be due to tautomerization of the product after the release, instead of
protonation by Lys211 since their experiments were conducted with cephalosporins having large R2 groups.
%

As next, we investigated the 1,3--proton shift occurring from N$_3$ to C$_5$ in {\bf EI3}, i.e. 
{\bf ES} $\rightarrow$ {\bf EI1} $\rightarrow$ {\bf EI2} $\rightarrow$ {\bf EI3} $\rightarrow$ {\bf EI4$^\prime$} ({\bf Path 4}); see \fref{fig:mechanism:all}.
In that simulation, we explicitly sampled  the tautomerization of double bond in the six membered ring of the drug molecule, however, no 1,3--proton
shift was observed even after applying a bias potential of 30~kcal~mol$^{-1}$ (SI~Figure~S18).
Thus, we excluded the possibility that {\bf Path 4} is a preferred reaction pathway.

Another pathway which we studied was {\bf Path 5}, i.e. {\bf ES} $\rightarrow$ {\bf EI1} $\rightarrow$ {\bf EI5}, where proton transfer 
from the protonated Lys211 occurs to N$_3$ (instead of C$_5$ as in {\bf Path 3}) , through W3 and W4 water molecules; see \fref{fig:mechanism:all}.
However, this reaction was not observed, even after applying a bias potential of 30~kcal~mol$^{-1}$ (SI~Figure~S19). 
Thus we conclude that {\bf Path 3} is the preferred protonation pathway compared to {\bf Path 5}.

As next we revisited meropenem hydrolysis mechanism which we carried in our previous work.~\cite{Ravi_2015_ACSCAT}
In particular, we performed computations of all elementary steps along {\bf Path 1} and {\bf Path 3} using the same procedures 
used for studying cephalexin hydrolysis, as before. see \fref{fig:FEProfile} and SI~Section~17.
Along {\bf Path 1} the free energy profile has qualitatively the same feature as that for cephalexin.
The free energy barrier for {\bf ES}$\rightarrow${\bf EI1} along {\bf Path 1} was also computed to be identical (20~kcal~mol$^{-1}$) for  both cephalexin and meropenem.
However, in the {\bf EI1} structure corresponding to meropenem, 
a water chain connecting Lys211 to C$_5$ was absent unlike in the case of cephalexin.
Thus, we exclude the possibility of {\bf Path 3} in the case of meropenem.
The origin of this difference between meropenem and cephalexin could be ascribed to the
differences in the {R2} groups of these drug molecules.
Clearly, cephalexin has a small {R2} group compared to that for meropenem; see \fref{general-mech-drugs}.
Our analysis indicates that the bulky {R2} group of meropenem could result in large steric effect 
such that it blocks the formation of stable water chain connecting Lys211 and C$_5$.
This is evident from \fref{int:gr}(a), where the number of water molecules surrounding the allylic C$_5$ atom is plotted, obtained by integrating the corresponding radial distribution function computed along the canonical ensemble trajectory of the {\bf EI1} structure.
In the case of cephalexin, there are two water molecules within 4~{\AA}  of C$_5$, while this is nearly zero for meropenem.
The  overlapped structures of {\bf EI1} with meropenem and cephlaxin in \fref{int:gr}(b), 
also indicates that position of W3 in cephalexin is occupied by the five membered ring part of the {R2} group of meropenem.
Thus, our study leads to the conclusion that depending on the bulkiness of the {R2} group, the protonation mechanism could change.

%
We compared the equilibrated structures of  the intermediates observed along {\bf Path 1} and {\bf Path 3} with the available X-ray structures; see SI~Section~9.
Strikingly, the crystal structure of NDM-1 with ring--opened cephalexin (PDB ID 4RL2~\cite{Feng_2014_JACS}) has
significant similarity with the {\bf EI2$^\prime$} structure; see SI~Figure~S7 and SI~Table~S5.
In particular, we find that the orientation of the proton at the C$_5$ position is identical with the crystal structure.
Thus, the crystal structure is more likely to be {\bf EI3$^\prime$} state, where Zn1 has a bound water molecule (via  {\bf EI2$^\prime$} $\rightarrow ${\bf EI3$^\prime$}).
This is a strong support for the proposed mechanism of protonation of C$_5$ from Lys211.
Additionally, the intermediate {\bf EI1} found in our study along the meropenem hydrolysis pathway matches 
very well with the crystal structure 
PDB ID 4EYL;~\cite{Dustin_2012_JACS} see SI~Figure~S7 and SI~Table~S6.
It is worth noting that no water molecules are bound to Zn1 and Zn2 in this crystal structure, thus confirming
our proposed mechanism that the coordination of a bulk water molecule occurs after the formation of {\bf EI1}. 
The crystal structure of NDM-1 with ampicillin (PDB ID: 3Q6X~\cite{Zhang_2011_FASEB}) having Zn1 bound water is 
likely to represent either {\bf EI2} or {\bf EI3}.
Such excellent agreements with the crystal structural data support the proposed reaction pathways in our study.
%

%
If reaction pathways {\bf Path 1} and {\bf Path 3} is casted to the following enzymatic reaction model  \\
\begin{center}
\setarrowdefault{0,0.7,black,thick}
\schemestart
 {\bf E} + {\bf S} \arrow{<=>[$k_1$][$k_{-1}$]}   {\bf ES} \arrow{->[$k_2$]}  {\bf EI1}   \arrow{->[$k_3$]} {\bf E} + {\bf P}
\schemestop
\end{center}
then
\begin{eqnarray} 
k_{\rm cat} = \frac{k_{2} k_{3} }{ k_{2} + k_{3} } \enspace .
\end{eqnarray}
In this model, we consider that $k_3$ is the effective rate constant for the decay of {\bf EI1}, which includes 
proton transfer to the substrate, water coordination to the active site, and product decoordination. 
Since ${\bf EI1} \rightarrow {\bf E}+{\bf P}$ involves water coordination from the bulk, and the barrier for this process is higher than that for the other elementary steps,  
we arrive at $k_{\rm cat} \approx k_{3}$.
%
%
In the kinetic experiments by Yang {\em et al.}~\cite{Yang_2012_BIOCHEMISTRY,Yang_2014_JACS}, the authors have found that $k_{\rm cat} \approx k_3$ for 
NDM-1 catalyzed
hydrolysis of nitrocefin and chromacef molecules.
%
%
Thus the experimental data is in line with our studies.
Further, both their experiment and our proposed mechanism supports the accumulation of anionic intermediates.
$k_{\rm cat}$ values measured for various cephalosporins and carbapenems catalyzed by NDM-1 and BcII enzymes do not vary much (which is only an order of magnitude).\cite{Yong_2009_AAC,Vila:PNAS:2005} 
This is consistent with our conclusion that $k_{\rm cat} \approx k_{\rm 3}$.
We also predict that solvent isotope effects will influence $k_{\rm cat}$ as {\bf EI1}$\rightarrow${\bf E}+{\bf P} involves water coordination and deprotonation.
This is consistent with the observation of solvent isotope effects in the experiments by Vila and co--workers.~\cite{Llarrull_2007_JBC} 

\section{Conclusion}
We find that the detailed mechanism of hydrolyisis by the two
drug molecules, namely cephalexin and meropenem, are different.
The ring--opening mechanism and the corresponding free energy barriers
are identical for both the drug molecules.
However, the mechanism of proton transfer steps for the two drug molecules 
differ. 
Protonation of the ring--opened substrate could occur in two different ways.
In one case, proton transfer occurs to the $\beta$--lactam nitrogen of the
ring--opened intermediate by a water molecule which diffuses into the active 
site after the ring--opening reaction.
Along the other route, proton transfer occurs from Lys211 to C$_5$ through a 
 chain of two water molecules.
The latter mechanism would take place when the drug molecules contain small 
{R2} group as in the case of cephalexin, while the former mechanism is active for drug 
molecules such as meropenem having a larger {R2} group.
Observation of different protonation pathways also agree well with the crystal 
structures of the reaction intermediates involving cephalexin and meropenem.
We conclude that $k_{\rm cat}\approx k_3$ and thus is determined mostly by the water coordination reaction
that occurs prior to the N$_3$ protonation in the case of nitrocefin, while it takes place after the C$_5$ protonation
in the case of cephalexin.
This is also in agreement with various experimental kinetic studies.
%
%
The detailed mechanistic picture, especially the structural and energetic 
data, presented in this work brings new molecular understanding on the catalytic reactions by 
NDM-1 and we believe that these results will aid in the ongoing research towards developing novel 
drugs and inhibitors targeting NDM-1.

\begin{acknowledgement}
Authors are grateful to IIT Kanpur for availing the HPC facility and Department of Biotechnology, India,
for funding this project. Authors
also acknowledge the technical discussions with Dr. Ravi Tripathi and Ms. Shalini Awasthi.
CKD thanks IIT Kanpur for his Ph.D. scholarship. 
%

\end{acknowledgement}


%




\begin{mcitethebibliography}{69}
\providecommand*\natexlab[1]{#1}
\providecommand*\mciteSetBstSublistMode[1]{}
\providecommand*\mciteSetBstMaxWidthForm[2]{}
\providecommand*\mciteBstWouldAddEndPuncttrue
  {\def\EndOfBibitem{\unskip.}}
\providecommand*\mciteBstWouldAddEndPunctfalse
  {\let\EndOfBibitem\relax}
\providecommand*\mciteSetBstMidEndSepPunct[3]{}
\providecommand*\mciteSetBstSublistLabelBeginEnd[3]{}
\providecommand*\EndOfBibitem{}
\mciteSetBstSublistMode{f}
\mciteSetBstMaxWidthForm{subitem}{(\alph{mcitesubitemcount})}
\mciteSetBstSublistLabelBeginEnd
  {\mcitemaxwidthsubitemform\space}
  {\relax}
  {\relax}

\bibitem[WHO(2014)]{WHO_REPORT_2014}
WHO, \emph{Antimicrobial resistance: global report on surveillance 2014}; WHO
  Press, World Health Organization: Geneva, 2014;
  http://www.who.int/drugresistance/documents/surveillancereport/en/ (accessed
  Oct 25, 2016)\relax
\mciteBstWouldAddEndPuncttrue
\mciteSetBstMidEndSepPunct{\mcitedefaultmidpunct}
{\mcitedefaultendpunct}{\mcitedefaultseppunct}\relax
\EndOfBibitem
\bibitem[AMR()]{AMR_REPORT_2016}
The Review on antimicrobial resistance. Final report. http://amr-review.
  org/Publications (accessed Oct. 25, 2016).\relax
\mciteBstWouldAddEndPunctfalse
\mciteSetBstMidEndSepPunct{\mcitedefaultmidpunct}
{}{\mcitedefaultseppunct}\relax
\EndOfBibitem
\bibitem[Yong \latin{et~al.}(2009)Yong, Toleman, Giske, Cho, Sundman, Lee, and
  Walsh]{Yong_2009_AAC}
Yong,~D.; Toleman,~M.~A.; Giske,~C.~G.; Cho,~H.~S.; Sundman,~K.; Lee,~K.;
  Walsh,~T.~R. Characterization of a new metallo--$\beta$--lactamase gene,
  blaNDM-1, and a novel Erythromycin esterase gene carried on a unique genetic
  structure in Klebsiella pneumoniae sequence type 14 from India.
  \emph{Antimicrob.~Agents~Chemother.} \textbf{2009}, \emph{53},
  5046--5054\relax
\mciteBstWouldAddEndPuncttrue
\mciteSetBstMidEndSepPunct{\mcitedefaultmidpunct}
{\mcitedefaultendpunct}{\mcitedefaultseppunct}\relax
\EndOfBibitem
\bibitem[Kumarasamy \latin{et~al.}(2010)Kumarasamy, Toleman, Walsh, Bagaria,
  Butt, Balakrishnan, Chaudhary, Doumith, Giske, Irfan, Krishnan, Kumar,
  Maharjan, Mushtaq, Noorie, Paterson, Pearson, Perry, Pike, Rao, Ray, Sarma,
  Sharma, Sheridan, Thirunarayan, Turton, Upadhyay, Warner, Welfare, Livermore,
  and Woodford]{Kumarasamy_2010_LID}
Kumarasamy,~K.~K.; Toleman,~M.~A.; Walsh,~T.~R.; Bagaria,~J.; Butt,~F.;
  Balakrishnan,~R.; Chaudhary,~U.; Doumith,~M.; Giske,~C.~G.; Irfan,~S.
  \latin{et~al.}  Emergence of a new antibiotic resistance mechanism in India,
  Pakistan, and the UK: a molecular, biological, and epidemiological study.
  \emph{Lancet~Infect.~Dis.} \textbf{2010}, \emph{10}, 597--602\relax
\mciteBstWouldAddEndPuncttrue
\mciteSetBstMidEndSepPunct{\mcitedefaultmidpunct}
{\mcitedefaultendpunct}{\mcitedefaultseppunct}\relax
\EndOfBibitem
\bibitem[Nordmann \latin{et~al.}(2011)Nordmann, Poirel, Walsh, and
  Livermore]{Nordmann_2011_TM}
Nordmann,~P.; Poirel,~L.; Walsh,~T.~R.; Livermore,~D.~M. The emerging NDM
  carbapenemases. \emph{Trends~Microbiol.} \textbf{2011}, \emph{19},
  588--595\relax
\mciteBstWouldAddEndPuncttrue
\mciteSetBstMidEndSepPunct{\mcitedefaultmidpunct}
{\mcitedefaultendpunct}{\mcitedefaultseppunct}\relax
\EndOfBibitem
\bibitem[Bushnell \latin{et~al.}(2013)Bushnell, Mitrani-Gold, and
  Mundy]{Mundy:2013}
Bushnell,~G.; Mitrani-Gold,~F.; Mundy,~L.~M. Emergence of New Delhi
  metallo--$\beta$--lactamase type 1-producing Enterobacteriaceae and
  non-Enterobacteriaceae: global case detection and bacterial surveillance.
  \emph{Int.~J.~Infect.~Dis} \textbf{2013}, \emph{17}, e325--e333\relax
\mciteBstWouldAddEndPuncttrue
\mciteSetBstMidEndSepPunct{\mcitedefaultmidpunct}
{\mcitedefaultendpunct}{\mcitedefaultseppunct}\relax
\EndOfBibitem
\bibitem[Tada \latin{et~al.}(2014)Tada, Shrestha, Miyoshi-Akiyama, Shimada,
  Ohara, Kirikae, and Pokhrel]{Tada_2014_AAC}
Tada,~T.; Shrestha,~B.; Miyoshi-Akiyama,~T.; Shimada,~K.; Ohara,~H.;
  Kirikae,~T.; Pokhrel,~B.~M. NDM-12, a novel New Delhi
  metallo--$\beta$--lactamase variant from a carbapenem-resistant escherichia
  coli clinical isolate in Nepal. \emph{Antimicrob.~Agents~Chemother.}
  \textbf{2014}, \emph{58}, 6302--6305\relax
\mciteBstWouldAddEndPuncttrue
\mciteSetBstMidEndSepPunct{\mcitedefaultmidpunct}
{\mcitedefaultendpunct}{\mcitedefaultseppunct}\relax
\EndOfBibitem
\bibitem[Shrestha \latin{et~al.}(2015)Shrestha, Tada, Miyoshi-Akiyama, Shimada,
  Ohara, Kirikae, and Pokhrel]{Shrestha_2015_AAC}
Shrestha,~B.; Tada,~T.; Miyoshi-Akiyama,~T.; Shimada,~K.; Ohara,~H.;
  Kirikae,~T.; Pokhrel,~B.~M. Identification of a novel NDM variant, NDM-13,
  from a multidrug--resistant Escherichia coli clinical isolate in Nepal.
  \emph{Antimicrob.~Agents~Chemother.} \textbf{2015}, \emph{59},
  5847--5850\relax
\mciteBstWouldAddEndPuncttrue
\mciteSetBstMidEndSepPunct{\mcitedefaultmidpunct}
{\mcitedefaultendpunct}{\mcitedefaultseppunct}\relax
\EndOfBibitem
\bibitem[Feng \latin{et~al.}(2014)Feng, Ding, Zhu, Liu, Xu, Zhang, Zang, Wang,
  and Liu]{Feng_2014_JACS}
Feng,~H.; Ding,~J.; Zhu,~D.; Liu,~X.; Xu,~X.; Zhang,~Y.; Zang,~S.; Wang,~D.;
  Liu,~W. Structural and mechanistic insights into NDM-1 catalyzed hydrolysis
  of cephalosporins. \emph{J.~Am.~Chem.~Soc.} \textbf{2014}, \emph{136},
  14694--14697\relax
\mciteBstWouldAddEndPuncttrue
\mciteSetBstMidEndSepPunct{\mcitedefaultmidpunct}
{\mcitedefaultendpunct}{\mcitedefaultseppunct}\relax
\EndOfBibitem
\bibitem[Zhang and Hao(2011)Zhang, and Hao]{Zhang_2011_FASEB}
Zhang,~H.; Hao,~Q. Crystal structure of NDM-1 reveals a common $\beta$--lactam
  hydrolysis mechanism. \emph{FASEB~J.} \textbf{2011}, \emph{25},
  2574--2582\relax
\mciteBstWouldAddEndPuncttrue
\mciteSetBstMidEndSepPunct{\mcitedefaultmidpunct}
{\mcitedefaultendpunct}{\mcitedefaultseppunct}\relax
\EndOfBibitem
\bibitem[Kim \latin{et~al.}(2013)Kim, Tesar, Jedrzejczak, Babnigg, Mire,
  Sacchettini, and Joachimiak]{Kim_2013_FASEB}
Kim,~Y.; Tesar,~C.; Jedrzejczak,~R.; Babnigg,~J.; Mire,~J.; Sacchettini,~J.;
  Joachimiak,~A. Structure of apo- and monometalated forms of NDM-1 -- a highly
  potent carbapenem-hydrolyzing metallo--$\beta$--lactamase. \emph{FASEB~J.}
  \textbf{2013}, \emph{27}, 1917--1927\relax
\mciteBstWouldAddEndPuncttrue
\mciteSetBstMidEndSepPunct{\mcitedefaultmidpunct}
{\mcitedefaultendpunct}{\mcitedefaultseppunct}\relax
\EndOfBibitem
\bibitem[King \latin{et~al.}(2012)King, Worrall, Gruninger, and
  Strynadka]{Dustin_2012_JACS}
King,~D.~T.; Worrall,~L.~J.; Gruninger,~R.; Strynadka,~N. C.~J. New Delhi
  metallo--$\beta$--lactamase: Structural insights into $\beta$--lactam
  recognition and inhibition. \emph{J.~Am.~Chem.~Soc.} \textbf{2012},
  \emph{134}, 11362--11365\relax
\mciteBstWouldAddEndPuncttrue
\mciteSetBstMidEndSepPunct{\mcitedefaultmidpunct}
{\mcitedefaultendpunct}{\mcitedefaultseppunct}\relax
\EndOfBibitem
\bibitem[Wang \latin{et~al.}(1998)Wang, Fast, , and Benkovic]{Wang_1998_JACS}
Wang,~Z.; Fast,~W.; ; Benkovic,~S.~J. Direct observation of an enzyme--bound
  intermediate in the catalytic cycle of the metallo--$\beta$--lactamase from
  Bacteroides fragilis. \emph{J.~Am.~Chem.~Soc.} \textbf{1998}, \emph{120},
  10788--10789\relax
\mciteBstWouldAddEndPuncttrue
\mciteSetBstMidEndSepPunct{\mcitedefaultmidpunct}
{\mcitedefaultendpunct}{\mcitedefaultseppunct}\relax
\EndOfBibitem
\bibitem[Rydzik \latin{et~al.}(2014)Rydzik, Brem, van Berkel, Pfeffer, Makena,
  Claridge, and Schofield]{Rydzik_2014_ANGEW}
Rydzik,~A.~M.; Brem,~J.; van Berkel,~S.~S.; Pfeffer,~I.; Makena,~A.;
  Claridge,~T. D.~W.; Schofield,~C.~J. Monitoring conformational changes in the
  NDM-1 metallo--$\beta$--lactamase by 19F NMR spectroscopy. \emph{Angew.~Chem.
  Int.~Ed.} \textbf{2014}, \emph{53}, 3129--3133\relax
\mciteBstWouldAddEndPuncttrue
\mciteSetBstMidEndSepPunct{\mcitedefaultmidpunct}
{\mcitedefaultendpunct}{\mcitedefaultseppunct}\relax
\EndOfBibitem
\bibitem[Yamaguchi \latin{et~al.}(2005)Yamaguchi, Kuroki, Yasuzawa, Higashi,
  Jin, Kawanami, Yamagata, Arakawa, Goto, and Kurosaki]{Yamaguchi_2005_JBC}
Yamaguchi,~Y.; Kuroki,~T.; Yasuzawa,~H.; Higashi,~T.; Jin,~W.; Kawanami,~A.;
  Yamagata,~Y.; Arakawa,~Y.; Goto,~M.; Kurosaki,~H. Probing the role of
  Asp--120(81) of metallo--$\beta$--lactamase (IMP-1) by site--directed
  mutagenesis, kinetic studies, and x--ray crystallography.
  \emph{J.~Biol.~Chem.} \textbf{2005}, \emph{280}, 20824--20832\relax
\mciteBstWouldAddEndPuncttrue
\mciteSetBstMidEndSepPunct{\mcitedefaultmidpunct}
{\mcitedefaultendpunct}{\mcitedefaultseppunct}\relax
\EndOfBibitem
\bibitem[Llarrull \latin{et~al.}(2007)Llarrull, Fabiane, Kowalski, Bennett,
  Sutton, and Vila]{Llarrull_2007_JBC}
Llarrull,~L.~I.; Fabiane,~S.~M.; Kowalski,~J.~M.; Bennett,~B.; Sutton,~B.~J.;
  Vila,~A.~J. Asp--120 locates Zn2 for optimal metallo--$\beta$--lactamase
  activity. \emph{J.~Biol.~Chem.} \textbf{2007}, \emph{282}, 18276--18285\relax
\mciteBstWouldAddEndPuncttrue
\mciteSetBstMidEndSepPunct{\mcitedefaultmidpunct}
{\mcitedefaultendpunct}{\mcitedefaultseppunct}\relax
\EndOfBibitem
\bibitem[Rasia and Vila(2004)Rasia, and Vila]{Rasia_2004_JBC}
Rasia,~R.~M.; Vila,~A.~J. Structural determinants of substrate binding to
  Bacillus cereus metallo--$\beta$--lactamase. \emph{J.~Biol.~Chem.}
  \textbf{2004}, \emph{279}, 26046--26051\relax
\mciteBstWouldAddEndPuncttrue
\mciteSetBstMidEndSepPunct{\mcitedefaultmidpunct}
{\mcitedefaultendpunct}{\mcitedefaultseppunct}\relax
\EndOfBibitem
\bibitem[Tioni \latin{et~al.}(2008)Tioni, Llarrull, Poeylaut-Palena,
  Mart$\acute{i}$, Saggu, Periyannan, Mata, Bennett, Murgida, and
  Vila]{Mariana_2008_JACS}
Tioni,~M.~F.; Llarrull,~L.~I.; Poeylaut-Palena,~A.~A.; Mart$\acute{i}$,~M.~A.;
  Saggu,~M.; Periyannan,~G.~R.; Mata,~E.~G.; Bennett,~B.; Murgida,~D.~H.;
  Vila,~A.~J. Trapping and characterization of a reaction intermediate in
  carbapenem hydrolysis by B. cereus metallo--$\beta$--lactamase.
  \emph{J.~Am.~Chem.~Soc.} \textbf{2008}, \emph{130}, 15852--15863\relax
\mciteBstWouldAddEndPuncttrue
\mciteSetBstMidEndSepPunct{\mcitedefaultmidpunct}
{\mcitedefaultendpunct}{\mcitedefaultseppunct}\relax
\EndOfBibitem
\bibitem[Breece \latin{et~al.}(2009)Breece, Hu, Bennett, Crowder, and
  Tierney]{Breece_2009_JACS}
Breece,~R.~M.; Hu,~Z.; Bennett,~B.; Crowder,~M.~W.; Tierney,~D.~L. Motion of
  the zinc Ions in catalysis by a dizinc metallo--$\beta$--lactamase.
  \emph{J.~Am.~Chem.~Soc.} \textbf{2009}, \emph{131}, 11642--11643\relax
\mciteBstWouldAddEndPuncttrue
\mciteSetBstMidEndSepPunct{\mcitedefaultmidpunct}
{\mcitedefaultendpunct}{\mcitedefaultseppunct}\relax
\EndOfBibitem
\bibitem[Hawk \latin{et~al.}(2009)Hawk, Breece, Hajdin, Bender, Hu, Costello,
  Bennett, Tierney, and Crowder]{Hawk_2009_JACS}
Hawk,~M.~J.; Breece,~R.~M.; Hajdin,~C.~E.; Bender,~K.~M.; Hu,~Z.;
  Costello,~A.~L.; Bennett,~B.; Tierney,~D.~L.; Crowder,~M.~W. Differential
  binding of Co(II) and Zn(II) to metallo--$\beta$--lactamase Bla2 from
  Bacillus anthracis. \emph{J.~Am.~Chem.~Soc.} \textbf{2009}, \emph{131},
  10753--10762\relax
\mciteBstWouldAddEndPuncttrue
\mciteSetBstMidEndSepPunct{\mcitedefaultmidpunct}
{\mcitedefaultendpunct}{\mcitedefaultseppunct}\relax
\EndOfBibitem
\bibitem[McManus-Munoz and Crowder(1999)McManus-Munoz, and
  Crowder]{Silvia_1999_BIOCHEMISTRY}
McManus-Munoz,~S.; Crowder,~M.~W. Kinetic mechanism of
  metallo--$\beta$--lactamase L1 from Stenotrophomonas maltophilia.
  \emph{Biochemistry} \textbf{1999}, \emph{38}, 1547--1553\relax
\mciteBstWouldAddEndPuncttrue
\mciteSetBstMidEndSepPunct{\mcitedefaultmidpunct}
{\mcitedefaultendpunct}{\mcitedefaultseppunct}\relax
\EndOfBibitem
\bibitem[Hu \latin{et~al.}(2008)Hu, Periyannan, Bennett, and
  Crowder]{Zhenxin_2008_JACS}
Hu,~Z.; Periyannan,~G.; Bennett,~B.; Crowder,~M.~W. Role of the Zn1 and Zn2
  sites in Metallo--$\beta$--lactamase L1. \emph{J.~Am.~Chem.~Soc.}
  \textbf{2008}, \emph{130}, 14207--14216\relax
\mciteBstWouldAddEndPuncttrue
\mciteSetBstMidEndSepPunct{\mcitedefaultmidpunct}
{\mcitedefaultendpunct}{\mcitedefaultseppunct}\relax
\EndOfBibitem
\bibitem[Yang \latin{et~al.}(2012)Yang, Aitha, Hetrick, Richmond, Tierney, and
  Crowder]{Yang_2012_BIOCHEMISTRY}
Yang,~H.; Aitha,~M.; Hetrick,~A.~M.; Richmond,~T.~K.; Tierney,~D.~L.;
  Crowder,~M.~W. Mechanistic and spectroscopic studies of
  metallo--$\beta$--lactamase NDM-1. \emph{Biochemistry} \textbf{2012},
  \emph{51}, 3839--3847\relax
\mciteBstWouldAddEndPuncttrue
\mciteSetBstMidEndSepPunct{\mcitedefaultmidpunct}
{\mcitedefaultendpunct}{\mcitedefaultseppunct}\relax
\EndOfBibitem
\bibitem[Yang \latin{et~al.}(2014)Yang, Aitha, Marts, Hetrick, Bennett,
  Crowder, and Tierney]{Yang_2014_JACS}
Yang,~H.; Aitha,~M.; Marts,~A.~R.; Hetrick,~A.; Bennett,~B.; Crowder,~M.~W.;
  Tierney,~D.~L. Spectroscopic and mechanistic studies of heterodimetallic
  forms of metallo--$\beta$--lactamase NDM-1. \emph{J.~Am.~Chem.~Soc.}
  \textbf{2014}, \emph{136}, 7273--7285\relax
\mciteBstWouldAddEndPuncttrue
\mciteSetBstMidEndSepPunct{\mcitedefaultmidpunct}
{\mcitedefaultendpunct}{\mcitedefaultseppunct}\relax
\EndOfBibitem
\bibitem[Zheng and Xu(2013)Zheng, and Xu]{Zheng_2013_JPCB}
Zheng,~M.; Xu,~D. New Delhi metallo--$\beta$--lactamase I: Substrate binding
  and catalytic mechanism. \emph{J.~Phys.~Chem. B} \textbf{2013}, \emph{117},
  11596--11607\relax
\mciteBstWouldAddEndPuncttrue
\mciteSetBstMidEndSepPunct{\mcitedefaultmidpunct}
{\mcitedefaultendpunct}{\mcitedefaultseppunct}\relax
\EndOfBibitem
\bibitem[Peraro \latin{et~al.}(2007)Peraro, Vila, Carloni, , and
  Klein]{Matteo_2007_JACS}
Peraro,~M.~D.; Vila,~A.~J.; Carloni,~P.; ; Klein,~M.~L. Role of zinc content on
  the catalytic efficiency of B1 metallo $\beta$--lactamases.
  \emph{J.~Am.~Chem.~Soc.} \textbf{2007}, \emph{129}, 2808--2816\relax
\mciteBstWouldAddEndPuncttrue
\mciteSetBstMidEndSepPunct{\mcitedefaultmidpunct}
{\mcitedefaultendpunct}{\mcitedefaultseppunct}\relax
\EndOfBibitem
\bibitem[Park \latin{et~al.}(2005)Park, Brothers, , and
  Jr.]{Hwangseo_2005_JACS}
Park,~H.; Brothers,~E.~N.; ; Jr.,~K. M.~M. Hybrid QM/MM and DFT investigations
  of the catalytic mechanism and inhibition of the dinuclear zinc
  metallo--$\beta$--lactamase CcrA from Bacteroides fragilis.
  \emph{J.~Am.~Chem.~Soc.} \textbf{2005}, \emph{127}, 4232--4241\relax
\mciteBstWouldAddEndPuncttrue
\mciteSetBstMidEndSepPunct{\mcitedefaultmidpunct}
{\mcitedefaultendpunct}{\mcitedefaultseppunct}\relax
\EndOfBibitem
\bibitem[Xu \latin{et~al.}(2007)Xu, Guo, , and Cui]{Xu_2007_JACS}
Xu,~D.; Guo,~H.; ; Cui,~Q. Antibiotic deactivation by a dizinc
  $\beta$--lactamase: Mechanistic insights from QM/MM and DFT studies.
  \emph{J.~Am.~Chem.~Soc.} \textbf{2007}, \emph{129}, 10814--10822\relax
\mciteBstWouldAddEndPuncttrue
\mciteSetBstMidEndSepPunct{\mcitedefaultmidpunct}
{\mcitedefaultendpunct}{\mcitedefaultseppunct}\relax
\EndOfBibitem
\bibitem[Xu \latin{et~al.}(2007)Xu, , Guo, and Cui]{Xu_2007_JPCA}
Xu,~D.; ; Guo,~H.; Cui,~Q. Antibiotic binding to dizinc $\beta$--lactamase L1
  from Stenotrophomonas maltophilia: SCC-DFTB/CHARMM and DFT studies.
  \emph{J.~Phys.~Chem. A} \textbf{2007}, \emph{111}, 5630--5636\relax
\mciteBstWouldAddEndPuncttrue
\mciteSetBstMidEndSepPunct{\mcitedefaultmidpunct}
{\mcitedefaultendpunct}{\mcitedefaultseppunct}\relax
\EndOfBibitem
\bibitem[Zhu \latin{et~al.}(2013)Zhu, Lu, Liang, Kong, Ye, Jin, Geng, Chen,
  Zheng, Jiang, Li, and Luo]{Zhu_2013_JCAMD}
Zhu,~K.; Lu,~J.; Liang,~Z.; Kong,~X.; Ye,~F.; Jin,~L.; Geng,~H.; Chen,~Y.;
  Zheng,~M.; Jiang,~H. \latin{et~al.}  A quantum mechanics/molecular mechanics
  study on the hydrolysis mechanism of New Delhi
  metallo--$\beta$--lactamase--1. \emph{J.~Comput.-Aided~Mol.~Des.}
  \textbf{2013}, \emph{27}, 247--256\relax
\mciteBstWouldAddEndPuncttrue
\mciteSetBstMidEndSepPunct{\mcitedefaultmidpunct}
{\mcitedefaultendpunct}{\mcitedefaultseppunct}\relax
\EndOfBibitem
\bibitem[Su$\acute{a}$rez \latin{et~al.}(2002)Su$\acute{a}$rez, Brothers, , and
  Kenneth M.~Merz]{Dimas_2002_BIOCHEMISTRY}
Su$\acute{a}$rez,~D.; Brothers,~E.~N.; ; Kenneth M.~Merz,~J. Insights into the
  structure and dynamics of the dinuclear zinc $\beta$--lactamase site from
  bacteroides fragilis. \emph{Biochemistry} \textbf{2002}, \emph{41},
  6615--6630\relax
\mciteBstWouldAddEndPuncttrue
\mciteSetBstMidEndSepPunct{\mcitedefaultmidpunct}
{\mcitedefaultendpunct}{\mcitedefaultseppunct}\relax
\EndOfBibitem
\bibitem[Tripathi and Nair(2015)Tripathi, and Nair]{Ravi_2015_ACSCAT}
Tripathi,~R.; Nair,~N.~N. Mechanism of meropenem hydrolysis by New Delhi
  metallo $\beta$--lactamase. \emph{ACS~Catal.} \textbf{2015}, \emph{5},
  2577--2586\relax
\mciteBstWouldAddEndPuncttrue
\mciteSetBstMidEndSepPunct{\mcitedefaultmidpunct}
{\mcitedefaultendpunct}{\mcitedefaultseppunct}\relax
\EndOfBibitem
\bibitem[Meini \latin{et~al.}(2015)Meini, Llarrull, and Vila]{Meini_2015_FEBS}
Meini,~M.-R.; Llarrull,~L.~I.; Vila,~A.~J. Overcoming differences: The
  catalytic mechanism of metallo--$\beta$--lactamases. \emph{FEBS~Lett.}
  \textbf{2015}, \emph{589}, 3419--3432\relax
\mciteBstWouldAddEndPuncttrue
\mciteSetBstMidEndSepPunct{\mcitedefaultmidpunct}
{\mcitedefaultendpunct}{\mcitedefaultseppunct}\relax
\EndOfBibitem
\bibitem[Umayal \latin{et~al.}(2012)Umayal, Tamilselvi, and Mugesh]{Mugesh:12}
Umayal,~M.; Tamilselvi,~A.; Mugesh,~G. In \emph{Prog.~Inorg.~Chem.};
  Karlin,~K.~D., Ed.; John~Wiely \& Sons, Inc., 2012; Vol.~57; pp
  395--443\relax
\mciteBstWouldAddEndPuncttrue
\mciteSetBstMidEndSepPunct{\mcitedefaultmidpunct}
{\mcitedefaultendpunct}{\mcitedefaultseppunct}\relax
\EndOfBibitem
\bibitem[Palzkill(2013)]{Palzkill:13}
Palzkill,~T. Metallo--$\beta$--lactamase structure and function.
  \emph{Ann.~N.~Y.~Acad.~Sci.} \textbf{2013}, \emph{1277}, 91--104\relax
\mciteBstWouldAddEndPuncttrue
\mciteSetBstMidEndSepPunct{\mcitedefaultmidpunct}
{\mcitedefaultendpunct}{\mcitedefaultseppunct}\relax
\EndOfBibitem
\bibitem[Crisp \latin{et~al.}(2007)Crisp, Conners, Garrity, Carenbauer,
  Crowder, , and Spencer]{Jonathan_2007_BIOCHEMISTRY}
Crisp,~J.; Conners,~R.; Garrity,~J.~D.; Carenbauer,~A.~L.; Crowder,~M.~W.; ;
  Spencer,~J. Structural basis for the role of Asp--120 in
  metallo--$\beta$--lactamases. \emph{Biochemistry} \textbf{2007}, \emph{46},
  10664--10674\relax
\mciteBstWouldAddEndPuncttrue
\mciteSetBstMidEndSepPunct{\mcitedefaultmidpunct}
{\mcitedefaultendpunct}{\mcitedefaultseppunct}\relax
\EndOfBibitem
\bibitem[Garrity \latin{et~al.}(2004)Garrity, Carenbauer, Herron, and
  Crowder]{Crowder:2004}
Garrity,~J.~D.; Carenbauer,~A.~L.; Herron,~L.~R.; Crowder,~M.~W. Metal binding
  Asp--120 in metallo-$\beta$-lactamase L1 from Stenotrophomonas maltophillia
  plays a crucial role in catalysis. \emph{J.~Biol.~Chem.} \textbf{2004},
  \emph{279}, 920--927\relax
\mciteBstWouldAddEndPuncttrue
\mciteSetBstMidEndSepPunct{\mcitedefaultmidpunct}
{\mcitedefaultendpunct}{\mcitedefaultseppunct}\relax
\EndOfBibitem
\bibitem[Wang \latin{et~al.}(1999)Wang, Fast, , and
  Benkovic]{Wang_1999_Biochemistry}
Wang,~Z.; Fast,~W.; ; Benkovic,~S.~J. On the mechanism of the
  metallo--$\beta$--lactamase from Bacteroides fragili. \emph{Biochemistry}
  \textbf{1999}, \emph{38}, 10013--10023\relax
\mciteBstWouldAddEndPuncttrue
\mciteSetBstMidEndSepPunct{\mcitedefaultmidpunct}
{\mcitedefaultendpunct}{\mcitedefaultseppunct}\relax
\EndOfBibitem
\bibitem[Spencer \latin{et~al.}(2005)Spencer, Read, Sessions, Howell,
  Blackburn, and Gamblin]{Gamblin:2005}
Spencer,~J.; Read,~J.; Sessions,~R.~B.; Howell,~S.; Blackburn,~G.~M.;
  Gamblin,~S.~J. Antibiotic recognition by binuclear
  metallo-$\beta$--lactamases revealed by x--ray crystallograph.
  \emph{J.~Am.~Chem.~Soc.} \textbf{2005}, \emph{127}, 14439--14444\relax
\mciteBstWouldAddEndPuncttrue
\mciteSetBstMidEndSepPunct{\mcitedefaultmidpunct}
{\mcitedefaultendpunct}{\mcitedefaultseppunct}\relax
\EndOfBibitem
\bibitem[Garau \latin{et~al.}(2005)Garau, Bebrone, Anne, Galleni,
  J.-M.Fr{\'e}re, and Dideberg]{Dieberg:2005}
Garau,~G.; Bebrone,~C.; Anne,~C.; Galleni,~M.; J.-M.Fr{\'e}re,; Dideberg,~O. A
  metallo--$\beta$--lactamase enzyme in action: crystal structures of the
  monozinc carabpenemase CphA and its complex with biapenem.
  \emph{J.~Mol.~Biol.} \textbf{2005}, \emph{345}, 785--795\relax
\mciteBstWouldAddEndPuncttrue
\mciteSetBstMidEndSepPunct{\mcitedefaultmidpunct}
{\mcitedefaultendpunct}{\mcitedefaultseppunct}\relax
\EndOfBibitem
\bibitem[Bounaga \latin{et~al.}(1998)Bounaga, Laws, Galleni, and
  Page]{Bounaga:98}
Bounaga,~S.; Laws,~A.~P.; Galleni,~M.; Page,~M.~I. The mechanism of catalysis
  and the inhibition of the Bacillus cereus zinc--dependent $\beta$--lactamase.
  \emph{Biochem. J.} \textbf{1998}, \emph{331}, 703--711\relax
\mciteBstWouldAddEndPuncttrue
\mciteSetBstMidEndSepPunct{\mcitedefaultmidpunct}
{\mcitedefaultendpunct}{\mcitedefaultseppunct}\relax
\EndOfBibitem
\bibitem[Laio \latin{et~al.}(2002)Laio, VandeVondele, and
  Rothlisberger]{Laio_2002_JCP}
Laio,~A.; VandeVondele,~J.; Rothlisberger,~U. A Hamiltonian electrostatic
  coupling scheme for hybrid Car-Parrinello molecular dynamics simulations.
  \emph{J.~Chem.~Phys.} \textbf{2002}, \emph{116}, 6941--6947\relax
\mciteBstWouldAddEndPuncttrue
\mciteSetBstMidEndSepPunct{\mcitedefaultmidpunct}
{\mcitedefaultendpunct}{\mcitedefaultseppunct}\relax
\EndOfBibitem
\bibitem[Laio and Parrinello(2002)Laio, and Parrinello]{Laio_2002_PNAS}
Laio,~A.; Parrinello,~M. Escaping free-energy minima. \emph{Proc. Natl. Acad.
  Sci.} \textbf{2002}, \emph{99}, 12562--6\relax
\mciteBstWouldAddEndPuncttrue
\mciteSetBstMidEndSepPunct{\mcitedefaultmidpunct}
{\mcitedefaultendpunct}{\mcitedefaultseppunct}\relax
\EndOfBibitem
\bibitem[Iannuzzi \latin{et~al.}(2003)Iannuzzi, Laio, and
  Parrinello]{Laio_2003_PRL}
Iannuzzi,~M.; Laio,~A.; Parrinello,~M. Efficient exploration of reactive
  potential energy surfaces using Car-Parrinello molecular dynamics.
  \emph{Phys.~Rev.~Lett.} \textbf{2003}, \emph{90}, 238302\relax
\mciteBstWouldAddEndPuncttrue
\mciteSetBstMidEndSepPunct{\mcitedefaultmidpunct}
{\mcitedefaultendpunct}{\mcitedefaultseppunct}\relax
\EndOfBibitem
\bibitem[Awasthi \latin{et~al.}(2016)Awasthi, Kapil, and
  Nair]{Shalini_2016_JCC}
Awasthi,~S.; Kapil,~V.; Nair,~N.~N. Sampling free energy surfaces as slices by
  combining umbrella sampling and metadynamics. \emph{J.~Comp.~Chem.}
  \textbf{2016}, \emph{37}, 1413--1424\relax
\mciteBstWouldAddEndPuncttrue
\mciteSetBstMidEndSepPunct{\mcitedefaultmidpunct}
{\mcitedefaultendpunct}{\mcitedefaultseppunct}\relax
\EndOfBibitem
\bibitem[Dupradeau \latin{et~al.}(2010)Dupradeau, Pigache, Zaffran, Savineau,
  Lelong, Grivel, Lelong, Rosanski, and Cieplak]{redtools}
Dupradeau,~F.; Pigache,~A.; Zaffran,~T.; Savineau,~C.; Lelong,~R.; Grivel,~N.;
  Lelong,~D.; Rosanski,~W.; Cieplak,~P. The R.E.D. tools: advances in RESP and
  ESP charge derivation and force field library building.
  \emph{Phys.~Chem.~Chem.~Phys.} \textbf{2010}, \emph{12}, 7821--7839\relax
\mciteBstWouldAddEndPuncttrue
\mciteSetBstMidEndSepPunct{\mcitedefaultmidpunct}
{\mcitedefaultendpunct}{\mcitedefaultseppunct}\relax
\EndOfBibitem
\bibitem[Wang \latin{et~al.}(2004)Wang, Wolf, Caldwell, Kollman, and
  Case]{gaff}
Wang,~J.; Wolf,~R.~M.; Caldwell,~J.~W.; Kollman,~P.~A.; Case,~D.~A. Development
  and testing of a general amber force field. \emph{J.~Comp.~Chem.}
  \textbf{2004}, \emph{25}, 1157--1174\relax
\mciteBstWouldAddEndPuncttrue
\mciteSetBstMidEndSepPunct{\mcitedefaultmidpunct}
{\mcitedefaultendpunct}{\mcitedefaultseppunct}\relax
\EndOfBibitem
\bibitem[Tripathi(August 2014)]{Ravi_thesis}
Tripathi,~R. \emph{Molecular mechanism of antibiotic resistance by class--C and
  New Delhi metallo $\beta$--lactamases: A QM/MM molecular dynamics study};
  Doctoral dissertation, IIT Kanpur, Kanpur, India, August 2014; see also {\tt
  http://172.28.64.70:8080/jspui/handle/123456789/14708} (accessed Oct 25,
  2016)\relax
\mciteBstWouldAddEndPuncttrue
\mciteSetBstMidEndSepPunct{\mcitedefaultmidpunct}
{\mcitedefaultendpunct}{\mcitedefaultseppunct}\relax
\EndOfBibitem
\bibitem[Cheatham \latin{et~al.}(1999)Cheatham, Cieplak, and Kollman]{parm99}
Cheatham,~T.~E.; Cieplak,~P.; Kollman,~P.~A. A modified version of the Cornell
  et al. force field with improved sugar pucker phases and helical repeat.
  \emph{J.~Biomol.~Struct.~Dyn.} \textbf{1999}, \emph{16}, 845--862\relax
\mciteBstWouldAddEndPuncttrue
\mciteSetBstMidEndSepPunct{\mcitedefaultmidpunct}
{\mcitedefaultendpunct}{\mcitedefaultseppunct}\relax
\EndOfBibitem
\bibitem[Jorgensen \latin{et~al.}(1983)Jorgensen, Chandrasekhar, Madura, Impey,
  and Klein]{tip3p}
Jorgensen,~W.~L.; Chandrasekhar,~J.; Madura,~J.~D.; Impey,~R.~W.; Klein,~M.~L.
  Comparison of simple potential functions for simulating liquid water.
  \emph{J.~Chem.~Phys.} \textbf{1983}, \emph{79}, 926--935\relax
\mciteBstWouldAddEndPuncttrue
\mciteSetBstMidEndSepPunct{\mcitedefaultmidpunct}
{\mcitedefaultendpunct}{\mcitedefaultseppunct}\relax
\EndOfBibitem
\bibitem[Darden \latin{et~al.}(1993)Darden, York, and Pedersen]{pme}
Darden,~T.; York,~D.; Pedersen,~L. Particle mesh Ewald: An Nlog(N) method for
  Ewald sums in large systems. \emph{J.~Phys.~Chem.} \textbf{1993}, \emph{98},
  10089--10092\relax
\mciteBstWouldAddEndPuncttrue
\mciteSetBstMidEndSepPunct{\mcitedefaultmidpunct}
{\mcitedefaultendpunct}{\mcitedefaultseppunct}\relax
\EndOfBibitem
\bibitem[Case \latin{et~al.}(2005)Case, Cheatham, Darden, Gohlke, Luo, Merz,
  Onufriev, Simmerling, Wang, and Woods]{amber}
Case,~D.~A.; Cheatham,~T.~E.; Darden,~T.; Gohlke,~H.; Luo,~R.; Merz,~K.~M.;
  Onufriev,~A.; Simmerling,~C.; Wang,~B.; Woods,~R.~J. The amber biomolecular
  simulation programs. \emph{J.~Comp.~Chem.} \textbf{2005}, \emph{26},
  1668--1688\relax
\mciteBstWouldAddEndPuncttrue
\mciteSetBstMidEndSepPunct{\mcitedefaultmidpunct}
{\mcitedefaultendpunct}{\mcitedefaultseppunct}\relax
\EndOfBibitem
\bibitem[Berendsen \latin{et~al.}(1984)Berendsen, Postma, van Gunsteren,
  DiNola, and Haak]{Berendsen_1984_JCP}
Berendsen,~H. J.~C.; Postma,~J. P.~M.; van Gunsteren,~W.~F.; DiNola,~A.;
  Haak,~J.~R. Molecular dynamics with coupling to an external bath.
  \emph{J.~Chem.~Phys.} \textbf{1984}, \emph{81}, 3684--3690\relax
\mciteBstWouldAddEndPuncttrue
\mciteSetBstMidEndSepPunct{\mcitedefaultmidpunct}
{\mcitedefaultendpunct}{\mcitedefaultseppunct}\relax
\EndOfBibitem
\bibitem[Loncharich \latin{et~al.}(1992)Loncharich, Brooks, and
  Pastor]{Loncharich_1992_BIOPOLYMERS}
Loncharich,~R.~J.; Brooks,~B.~R.; Pastor,~R.~W. Langevin dynamics of peptides:
  The frictional dependence of isomerization rates of
  N-acetylalanyl-N$^\prime$-methylamide. \emph{Biopolymers} \textbf{1992},
  \emph{32}, 523--535\relax
\mciteBstWouldAddEndPuncttrue
\mciteSetBstMidEndSepPunct{\mcitedefaultmidpunct}
{\mcitedefaultendpunct}{\mcitedefaultseppunct}\relax
\EndOfBibitem
\bibitem[Warshel and Levitt(1976)Warshel, and Levitt]{Warshel_1976_JMB}
Warshel,~A.; Levitt,~M. Theoretical studies of enzymic reactions: Dielectric,
  electrostatic and steric stabilization of the carbonium ion in the reaction
  of lysozyme. \emph{J.~Mol.~Biol.} \textbf{1976}, \emph{103}, 227 -- 249\relax
\mciteBstWouldAddEndPuncttrue
\mciteSetBstMidEndSepPunct{\mcitedefaultmidpunct}
{\mcitedefaultendpunct}{\mcitedefaultseppunct}\relax
\EndOfBibitem
\bibitem[Carloni \latin{et~al.}(2002)Carloni, Rothlisberger, and
  Parrinello]{Carloni:2002}
Carloni,~P.; Rothlisberger,~U.; Parrinello,~M. The role and perspective of ab
  initio molecular dynamics in the study of biological systems.
  \emph{Acc.~Chem.~Res.} \textbf{2002}, \emph{35}, 455--464\relax
\mciteBstWouldAddEndPuncttrue
\mciteSetBstMidEndSepPunct{\mcitedefaultmidpunct}
{\mcitedefaultendpunct}{\mcitedefaultseppunct}\relax
\EndOfBibitem
\bibitem[Marx and Hutter(2009)Marx, and Hutter]{Dominik_AIMD_book}
Marx,~D.; Hutter,~J. \emph{Ab Initio Molecular Dynamics}; Cambridge University
  Press, 2009\relax
\mciteBstWouldAddEndPuncttrue
\mciteSetBstMidEndSepPunct{\mcitedefaultmidpunct}
{\mcitedefaultendpunct}{\mcitedefaultseppunct}\relax
\EndOfBibitem
\bibitem[cpm()]{cpmdpackage}
{\tt CPMD}, J. Hutter et al., IBM Corp 1990-2004, MPI f{\"u}r
  Festk{\"o}rperforschung Stuttgart 1997-2001, see also {\tt
  http://www.cpmd.org}\relax
\mciteBstWouldAddEndPuncttrue
\mciteSetBstMidEndSepPunct{\mcitedefaultmidpunct}
{\mcitedefaultendpunct}{\mcitedefaultseppunct}\relax
\EndOfBibitem
\bibitem[Perdew \latin{et~al.}(1992)Perdew, Chevary, Vosko, Jackson, Pederson,
  Singh, and Fiolhais]{Perdew_1992_PRB}
Perdew,~J.~P.; Chevary,~J.~A.; Vosko,~S.~H.; Jackson,~K.~A.; Pederson,~M.~R.;
  Singh,~D.~J.; Fiolhais,~C. Atoms, molecules, solids, and surfaces:
  Applications of the generalized gradient approximation for exchange and
  correlation. \emph{Phys.~Rev.~B} \textbf{1992}, \emph{46}, 6671--6687\relax
\mciteBstWouldAddEndPuncttrue
\mciteSetBstMidEndSepPunct{\mcitedefaultmidpunct}
{\mcitedefaultendpunct}{\mcitedefaultseppunct}\relax
\EndOfBibitem
\bibitem[Vanderbilt(1990)]{Vanderbilt_1990_PRB}
Vanderbilt,~D. Soft self-consistent pseudopotentials in a generalized
  eigenvalue formalism. \emph{Phys.~Rev.~B} \textbf{1990}, \emph{41},
  7892--7895\relax
\mciteBstWouldAddEndPuncttrue
\mciteSetBstMidEndSepPunct{\mcitedefaultmidpunct}
{\mcitedefaultendpunct}{\mcitedefaultseppunct}\relax
\EndOfBibitem
\bibitem[Car and Parrinello(1985)Car, and Parrinello]{Car_1985_PRL}
Car,~R.; Parrinello,~M. Unified approach for molecular dynamics and
  density--functional theory. \emph{Phys.~Rev.~Lett.} \textbf{1985}, \emph{55},
  2471--2474\relax
\mciteBstWouldAddEndPuncttrue
\mciteSetBstMidEndSepPunct{\mcitedefaultmidpunct}
{\mcitedefaultendpunct}{\mcitedefaultseppunct}\relax
\EndOfBibitem
\bibitem[Martyna \latin{et~al.}(1992)Martyna, Klein, and
  Tuckerman]{Nose_1992_JCP}
Martyna,~G.~J.; Klein,~M.~L.; Tuckerman,~M. Nos{\'e}--Hoover chains: The
  canonical ensemble via continuous dynamics. \emph{J.~Chem.~Phys.}
  \textbf{1992}, \emph{97}, 2635--2643\relax
\mciteBstWouldAddEndPuncttrue
\mciteSetBstMidEndSepPunct{\mcitedefaultmidpunct}
{\mcitedefaultendpunct}{\mcitedefaultseppunct}\relax
\EndOfBibitem
\bibitem[Laio and Gervasio(2008)Laio, and Gervasio]{Laio_2008_RPP}
Laio,~A.; Gervasio,~F.~L. Metadynamics: a method to simulate rare events and
  reconstruct the free energy in biophysics, chemistry and material science.
  \emph{Rep.~Prog.~Phys.} \textbf{2008}, \emph{71}, 126601\relax
\mciteBstWouldAddEndPuncttrue
\mciteSetBstMidEndSepPunct{\mcitedefaultmidpunct}
{\mcitedefaultendpunct}{\mcitedefaultseppunct}\relax
\EndOfBibitem
\bibitem[Ensing \latin{et~al.}(2006)Ensing, Vivo, Liu, Moore, , and
  Klein]{Bernd_2006_ACR}
Ensing,~B.; Vivo,~M.~D.; Liu,~Z.; Moore,~P.; ; Klein,~M.~L. Metadynamics as a
  tool for exploring free energy landscapes of chemical reactions.
  \emph{Acc.~Chem.~Res.} \textbf{2006}, \emph{39}, 73--81\relax
\mciteBstWouldAddEndPuncttrue
\mciteSetBstMidEndSepPunct{\mcitedefaultmidpunct}
{\mcitedefaultendpunct}{\mcitedefaultseppunct}\relax
\EndOfBibitem
\bibitem[Barducci \latin{et~al.}(2011)Barducci, Bonomi, and
  Parrinello]{Barducci_2011_CMS}
Barducci,~A.; Bonomi,~M.; Parrinello,~M. Metadynamics.
  \emph{Wiley~Interdiscip.~Rev.~Comput.~Mol.~Sci.} \textbf{2011}, \emph{1},
  826--843\relax
\mciteBstWouldAddEndPuncttrue
\mciteSetBstMidEndSepPunct{\mcitedefaultmidpunct}
{\mcitedefaultendpunct}{\mcitedefaultseppunct}\relax
\EndOfBibitem
\bibitem[Valsson \latin{et~al.}(2016)Valsson, Tiwary, and
  Parrinello]{Valsson_2016_ARPC}
Valsson,~O.; Tiwary,~P.; Parrinello,~M. Enhancing important fluctuations: Rare
  events and metadynamics from a conceptual viewpoint.
  \emph{Annu.~Rev.~Phys.~Chem.} \textbf{2016}, \emph{67}, 159--184\relax
\mciteBstWouldAddEndPuncttrue
\mciteSetBstMidEndSepPunct{\mcitedefaultmidpunct}
{\mcitedefaultendpunct}{\mcitedefaultseppunct}\relax
\EndOfBibitem
\bibitem[Dellago \latin{et~al.}(1998)Dellago, Bolhuis, Csajka, and
  Chandler]{Christoph_TPS}
Dellago,~C.; Bolhuis,~P.~G.; Csajka,~F.~S.; Chandler,~D. Transition path
  sampling and the calculation of rate constants. \emph{J.~Chem.~Phys.}
  \textbf{1998}, \emph{108}\relax
\mciteBstWouldAddEndPuncttrue
\mciteSetBstMidEndSepPunct{\mcitedefaultmidpunct}
{\mcitedefaultendpunct}{\mcitedefaultseppunct}\relax
\EndOfBibitem
\bibitem[Tomatis \latin{et~al.}(2005)Tomatis, Rasia, Segovia, and
  Vila]{Vila:PNAS:2005}
Tomatis,~P.~E.; Rasia,~R.~M.; Segovia,~L.; Vila,~A.~J. Mimicking natural
  evolution in metallo--$\beta$--lactamases through second-shell ligand
  mutations. \emph{Proc. Natl. Acad. Sci.} \textbf{2005}, \emph{102},
  13761--13766\relax
\mciteBstWouldAddEndPuncttrue
\mciteSetBstMidEndSepPunct{\mcitedefaultmidpunct}
{\mcitedefaultendpunct}{\mcitedefaultseppunct}\relax
\EndOfBibitem
\end{mcitethebibliography}
\bibliographystyle{achemso}
\providecommand{\latin}[1]{#1}
\providecommand*\mcitethebibliography{\thebibliography}
\csname @ifundefined\endcsname{endmcitethebibliography}
  {\let\endmcitethebibliography\endthebibliography}{}

\end{document}